\documentclass[,final,sort&compress]{aipproc}

\layoutstyle{6x9}
   \usepackage{graphicx}
   \usepackage{epstopdf}
  \usepackage{amssymb}

\def\teq#1{$\, #1\,$}                         
\def\apj{ApJ}

\def\asr{Adv. Space Res.}                       
\def\jgr{J. Geophys. Res.}

\def\ssr{Space Sci. Rev.}                       
\newcommand{\vol}[2]{$\,$\rm #1\rm , #2.}           
             \font\sevenrm=cmr7

\def\thetascatt{\theta_{\hbox{\sevenrm scatt}}}
\def\pmin{p_{\hbox{\sevenrm min}}}
\def\pmax{p_{\hbox{\sevenrm max}}}
\def\dover#1#2{\hbox{${{\displaystyle#1 \vphantom{(} }\over{
   \displaystyle #2 \vphantom{(} }}$}}

\begin{document}
\begin{flushright}
\phantom{p}
\vspace{-60pt}
     To appear in Proc. of the {\it 7th IGPP International Astrophysics Conference}\\ 
     ``Particle Acceleration and Transport in the Heliosphere and Beyond'' (2008),\\
     eds. G. Li, et al. (AIP Conf. Proc., New York).
\end{flushright} 

\title{Particle Acceleration at Interplanetary Shocks}

\author{Matthew G. Baring \& Errol J. Summerlin}{
    address={Department of Physics and Astronomy, MS-108,
                      Rice University, P. O. Box 1892, \\
                      Houston, TX 77251-1892, USA\\
                      {\rm Email: baring@rice.edu, xerex@rice.edu}}
}



\begin{abstract}
The acceleration of interstellar pick-up ions as well as solar wind
species has been observed at a multitude of interplanetary (IP) shocks
by different spacecraft. The efficiency of injection of the pick-up ion
component differs from that of the solar wind, and is strongly enhanced
at highly oblique and quasi-perpendicular shock events. This paper
expands upon previous work modeling the phase space distributions of
accelerated ions associated with the shock event encountered on day 292
of 1991 by the Ulysses mission at 4.5 AU.  As in the prior work, a
kinetic Monte Carlo simulation is employed here to model the diffusive
acceleration process. This exposition presents recent developments
pertaining to the incorporation into the simulation of the diffusive
characteristics incurred by field line wandering (FLW), according to the
work of Giacalone and Jokipii. Resulting ion distributions and upstream
diffusion scales are presented and compared with Ulysses data.  For a
pure field-line wandering construct, it is determined that the upstream
spatial ramp scales are too short to accommodate the HI-SCALE flux
increases for 200 keV protons, and that the distribution function for
\teq{H^+}  somewhat underpopulates the combined SWICS/HI-SCALE spectra
at the shock.  This contrasts our earlier theory/data comparison where
it was demonstrated that diffusive transport in highly turbulent fields
according to kinetic theory can successfully account for both the proton
distribution data near the shock, and the observation of energetic
protons upstream of this interplanetary shock, using a single turbulence
parameter. The principal conclusion here is that, in a FLW scenario, the
transport of ions across the mean magnetic field is slightly less
efficient than is required to effectively trap energetic ions within a
few Larmor radii of the shock layer, and thereby sustain acceleration at
levels that match the observed distributions. This highlights the
contrast between ion transport in highly turbulent shock environs and
remote, less-disturbed interplanetary regions.
\end{abstract}

\maketitle

\section{Introduction}
\label{sec:Introduction}
There is bountiful evidence for efficient particle acceleration at
collisionless shocks in the heliosphere, and interplanetary (IP) shocks
provide interesting and useful test cases for shock acceleration
theories.  For observations of energetic ion populations at IP shocks in
the pre-Ulysses era, see, for example, \cite{SvA74,DPK81,Tan88}.  A
leading hypothesis is that in these systems, both solar wind and pick-up
ions are energized to above 200 keV by the mechanism of diffusive shock
acceleration. This process forms the focus of this paper, which employs
the kinematic Monte Carlo technique of Ellison and Jones (e.g.,
\cite{EJE81,JE91,EBJ95}) to model ion diffusion and convection in the
turbulent shock environs, and describes the acceleration of particles
that are injected directly from the thermal and pick-up ion populations.
Upstream plasma quantities are input from observational data, and
downstream quantities are determined using the full MHD Rankine-Hugoniot
relations. The Monte Carlo method was employed by Ellison et al.
\cite{EMP90} to perform the first successful theory/data comparison for
the quasi-parallel portion of the Earth's bow shock, using proton,
\teq{He^{++}} and other ion distributions from the AMPTE experiment.

A similar theory/data comparison was enunciated for IP shocks in Baring
et al. \cite{BOEF97}, where impressive agreement was found between the
simulation predictions and spectral data obtained by the Solar Wind Ion
Composition Spectrometer (SWICS) aboard Ulysses, in the case of two
highly-oblique shocks observed early in the Ulysses mission.  Such
agreement was possible only with the assumption of diffusion fairly near
the Bohm limit.  Summerlin \& Baring \cite{BS05,SB06} extended this
program to explore the role of pick-up ions in the acceleration process
at an IP shock detected by Ulysses' SWICS and HI-SCALE instruments at
around 4.5 AU, as reported in \cite{Gloeck94}.  Phase space
distributions from the simulations, in the case of large angle
scattering, were compared with the data, yielding acceptable fits for
the proton populations and upstream flux increases, using standard
prescriptions for the injected pick-up ion distribution. This paper
further extends our investigative program by exploring whether or not
diffusion due to field line wandering can model Ulysses' observations
for this shock equally well.

\section{Modeling Ulysses Data for the Day 292, 1991 Shock}
 \label{sec:model}

This work thus continues our case study of the forward shock of a CIR
encountered by Ulysses on Day 292 of 1991, for which downstream particle
distributions were published in Gloeckler et al. \cite{Gloeck94}.
Various plasma parameters for this oblique shock were input for the
Monte Carlo simulation, and were obtained from \cite{Gloeck94} and the
data compilations of \cite{Balogh95,Hoang95}.   Most important among
these was the angle \teq{\theta_{Bn1} = 50^{\circ}\pm 11^{\circ}} the
upstream magnetic field made with the shock normal. This shock was also
quite weak, with a sonic Mach number of  \teq{M_{\rm s}\sim 2.53}, and
an inferred value \cite{Gloeck94} of \teq{r=u_1/u_2=2.4 \pm 0.3} for the
velocity compression ratio. The normalization of solar wind proton
distributions was established using \teq{n_p=2.0}cm$^{-3}$ as the solar
wind proton density.  Other parameters used in the simulation, such as
the upstream flow speed of \teq{u_{1} \approx 55} km/s in the shock rest
frame, and upstream plasma temperatures, are detailed in Summerlin \&
Baring \cite{SB06}, as is the pick-up proton distribution input that was
taken from \cite{EJB99} (see also \cite{leRoux96}).

The Monte Carlo shock acceleration simulation is described in
\cite{JE91,EBJ95,BOEF97,SB06}.  Particles are injected upstream and
allowed to convect into the shock, meanwhile diffusing in space so as to
effect multiple shock crossings, and thereby gain energy through the
shock drift and Fermi processes.  The particles gyrate in laminar
electromagnetic fields, with their trajectories being obtained by
solving the Lorentz force equation in the shock rest frame, in which
there is, in general, a {\bf u $\times$ B} electric field in addition to
the magnetic field. The effects of magnetic turbulence are modeled by
phenomenologically scattering these ions elastically in the rest frame
of the local fluid flow. The simulation outputs particle fluxes and
phase space distributions at any location upstream or downstream of the
shock, and in any reference frame including that of the Ulysses
spacecraft.

The simulation can routinely model either large-angle or small-angle
scattering. At every scattering, the direction of the particle's
momentum vector is deflected in the local fluid frame within some solid
angle.  The resulting effect is that the gyrocenter of a particle is
shifted randomly by a distance of the order of one gyroradius or less in
the plane orthogonal to the local field.  Accordingly, cross-field
diffusion emerges naturally from the simulation.  For large angle
scattering (LAS), the scattering solid angle is \teq{4\pi} steradians,
and the transport is governed by kinetic theory  \cite{FJO74,EBJ95},
where the ratio of the spatial diffusion coefficients parallel
(\teq{\kappa_\parallel =\lambda v/3}) and perpendicular
(\teq{\kappa_\perp}) to the mean magnetic field is given by
\teq{\kappa_\perp /\kappa_\parallel = 1/(1+\eta^2)}.  Here, the
parameter \teq{\eta =\lambda/r_g} is the ratio of a particle's mean free
path \teq{\lambda} to its gyroradius \teq{r_g}.  Clearly, \teq{\eta}
controls the amount of cross-field diffusion, and is a measure of the
level of turbulence present in the system, i.e. is an indicator of
\teq{\langle \delta B/B\rangle}.  The Bohm limit of quasi-isotropic
diffusion is realized when \teq{\eta\sim 1} and \teq{\langle \delta
B/B\rangle\sim 1}.   In LAS applications, \teq{\eta} is prescribed to be
independent of momentum.  In general, it is a parameter that critically
controls the injection efficiency of low energy particles, and the
upstream diffusion scale of accelerated ions.

The Monte Carlo simulation can specify arbitrary ratios of perpendicular
to parallel diffusion \teq{\kappa_{\perp}/\kappa_{\parallel}} in a
variety of ways.  The implementation in this work employed a construct
that embodies the key signatures of the field-line wandering (FLW) study
of Giacalone and Jokipii \cite{GJ99}: they found that (i)
\teq{\kappa_{\perp}/\kappa_{\parallel}} does not vary significantly with
particle momentum, and (ii) \teq{\kappa_{\parallel}=\lambda_{\parallel}
v/3} is a weak function of momentum, namely
\teq{\kappa_{\parallel}\propto p^{1+\alpha}} for \teq{\alpha\sim 1/3}.
This second property is readily incorporated in the Monte Carlo
technique via a momentum-dependent parameter \teq{\eta =\eta (p)}, i.e.
\teq{\lambda_{\parallel} = \eta r_g\propto p^{\alpha} \Rightarrow \eta
(p)\propto p^{\alpha -1}}, so that \teq{\alpha\approx 1/3}, and
\teq{\eta (p)} declines with increasing momentum.  The approximate
constancy of the ratio \teq{\kappa_{\perp}/\kappa_{\parallel}} can then
be accommodated by adapting the small-angle scattering implementation
specified in \cite{EJR90}.  Particle momenta are deflected uniformly
within a cone of opening angle \teq{\thetascatt\ll 1}, whose axis is
centered on the pre-scattering vector momentum.  Note that
\teq{\kappa_{\perp}} and \teq{\kappa_{\parallel}} can be analytically
coupled to field turbulence models via quasi-linear theory. 
Approximately \teq{\pi/\thetascatt^2} deflections accumulate to effect
an isotropization of the momenta.  In a single scattering event with
\teq{\thetascatt\ll 1}, the effective diffusive mean free path across
the field is of the order of \teq{\lambda_{\perp}\sim 2\thetascatt\,
r_g/3}.  It then follows from kinetic theory that \teq{\kappa_\perp
/\kappa_\parallel \approx \thetascatt^2/[6(1+\eta^2)]}.  As \teq{\eta\gg
1} for most momenta, then \teq{\thetascatt\propto \eta\propto p^{\alpha
-1}} is required to achieve constancy of
\teq{\kappa_{\perp}/\kappa_{\parallel}}. Let the operating momentum
range of particles of interest for this IP shock acceleration problem be
\teq{\pmin \lesssim p\leq \pmax}. Two global parameters are then defined
for the simulation runs: \teq{\theta_0= \thetascatt (\pmin )} and
\teq{\eta_0 = \eta (\pmin )} for \teq{\pmin = m_pu_1}.  Hence
\begin{equation}
   \thetascatt (p)\; =\; \theta_0\, \biggl( \dover{p}{\pmin} \biggr)^{\alpha-1}
   \quad ,\quad 
   \eta (p)\; =\; \eta_0\, \biggl( \dover{p}{\pmin} \biggr)^{\alpha-1}\quad ,
   \quad \alpha\;\approx\; 1/3\quad ,
 \label{eq:theta_eta_momdep}
\end{equation}
then encapsulate the momentum dependences of the diffusion parallel to
and perpendicular to the mean field.  Field fluctuations in typical IP
shocks are tantamount to magnetic turbulence variances of
\teq{\sigma^2\sim 0.03-0.1 B^2}, corresponding to ratios
\teq{\kappa_\perp /\kappa_\parallel \sim 10^{-4}-10^{-3}} according to
the field line wandering results in Figure 4 of \cite{GJ99}.  This then
sets \teq{\theta_0/\eta_0\sim 10^{-2}-10^{-1}} for the FLW-initiated
spatial diffusion modeling here.  One final nuance must be identified. 
The cumulative effect of \teq{\sim \pi/\thetascatt^2} deflections does
not amount to \teq{\kappa_{\parallel} = \eta r_g v/3} {\it per se},
unless the vector momentum is more or less reversed on the
\teq{\lambda_{\parallel}} scale.  Hence, the implementation of
FLW-instigated spatial diffusion here imposes an approximate reversal of
the direction of the component of momentum parallel to {\bf B} every
scattering event.  Physically this amounts to a magnetic mirroring
event, as is implied by the FLW study of \cite{GJ99}, superposed on a
small amount of isotropic diffusion.


\begin{figure}
 \centerline{
  \includegraphics[width=.51\textwidth]{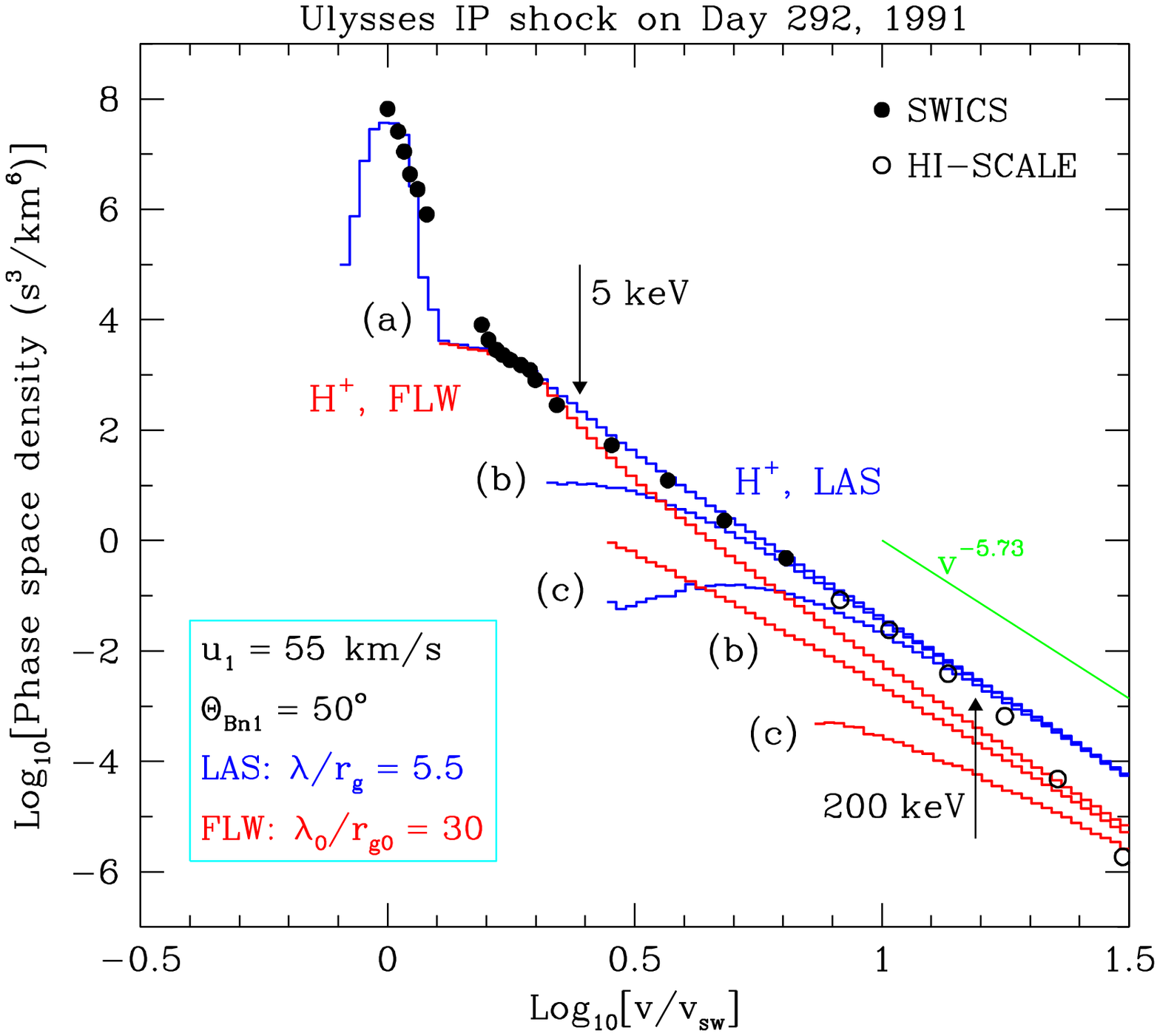}
   \hskip -0.1truecm
    \includegraphics[width=.51\textwidth]{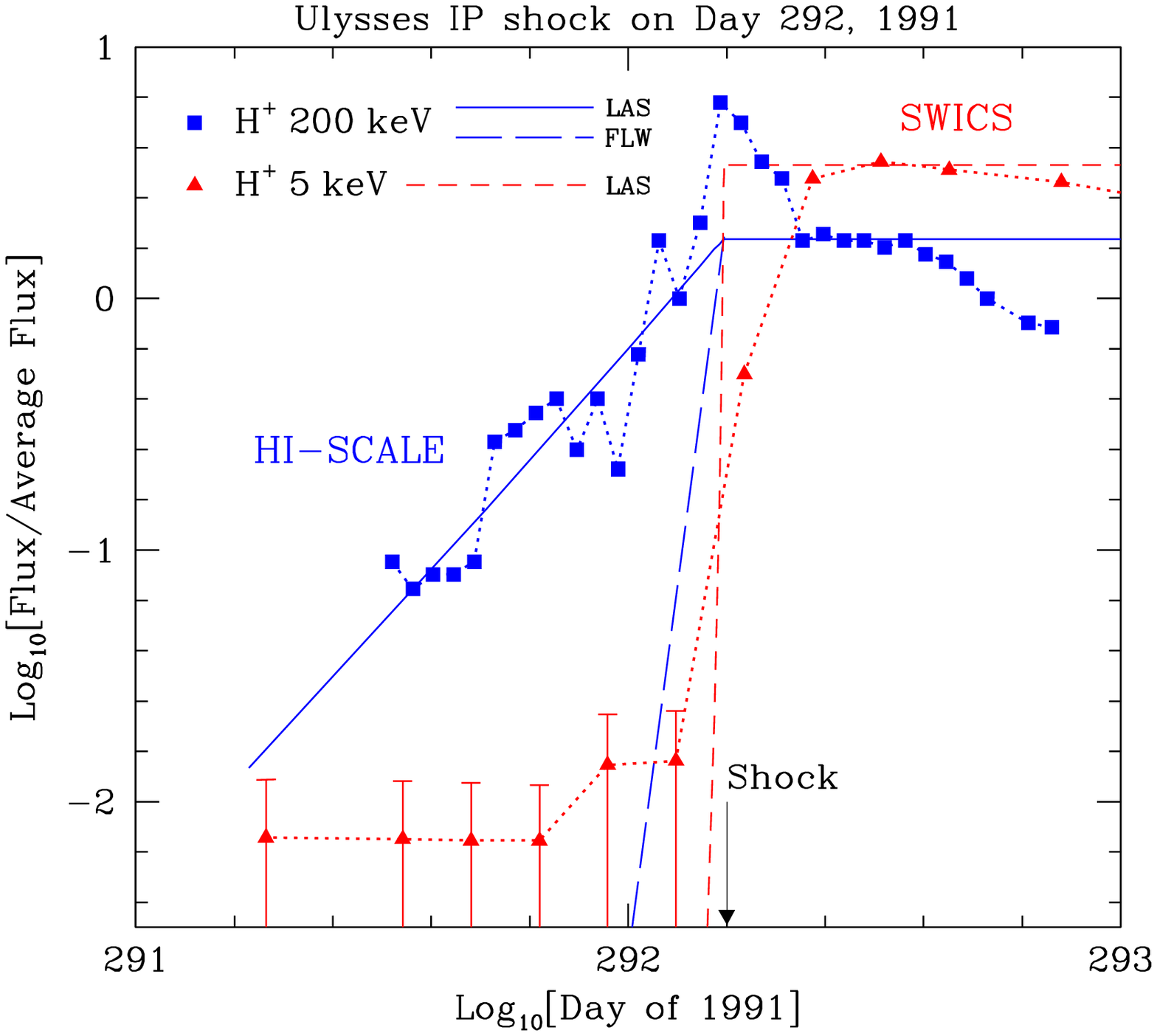}
  }
  \caption{{\it Left panel:} Comparison between phase space velocity 
distributions for downstream data collected by the Ulysses mission for
the shock on day 292 of 1991, and Monte Carlo model results.  The data
are for \teq{H^+}  solar wind and pickup ions (filled circles for SWICS
data; open circles for HI-SCALE points), and are taken from Gloeckler et
al. \cite{Gloeck94}.  The heavyweight histograms are the corresponding
Monte Carlo models of acceleration of protons for \teq{u_1=55}km/sec,
using plasma shock parameters from \cite{Gloeck94} and sources indicated
in the text (see also \cite{SB06}).  Two model cases are depicted -- the
blue (labeled \teq{H^+}, LAS) denotes a large angle scattering
implementation with \teq{\eta =\lambda /r_g=5.5}, and the red (labeled
\teq{H^+}, FLW) corresponds to a field line wandering diffusive model
with \teq{\eta_0\equiv\lambda_0/r_{g0}=30}. In each case, there are
three spectra shown, corresponding to (a) just downstream, and
successively increasing times upstream of the shock encounter, i.e. (b)
14 minutes and (c) 69 minutes.  The velocity axis is the ratio of the
ion speed \teq{v}, as measured in the spacecraft frame, to the solar
wind speed \teq{v_{sw}}.  The shock compression ratio was \teq{r=2.1},
implying diffusive acceleration power-laws of index \teq{-5.73},
indicated in green.
\newline
{\it Right panel:} The flux variations of accelerated pick-up ion
populations as a function of time near the shock.  The data for 5 keV
and 200 keV pick-up \teq{H^+} are depicted by filled red triangles and blue
squares, respectively, and are taken from \cite{Gloeck94}. The Monte
Carlo model generated fluxes at different distances normal to the shock,
which were converted to spacecraft times by incorporating solar wind
convection.  The 5 keV and 200 keV pick-up \teq{H^+} LAS model traces are
displayed as red dashed and blue solid curves, respectively, and exhibit an
exponential decline upstream of the shock that is characteristic of diffusive 
shock acceleration.  The 200 keV FLW model profile (blue, long dashes)
displays an upstream ramp that is much shorter than the observed profile, 
due to low \teq{\lambda /r_g} values at these high ion energies.
}
\end{figure}

\phantom{p}
\vskip -25pt
Results for the acceleration of solar wind and pick-up ions for the case
of large angle scattering in strong field turbulence (\teq{\langle\delta
B/B\rangle\gtrsim 0.1}) were presented and discussed at length in
\cite{BS05,SB06}.  A selection of these are exhibited in Figure~1, which
displays downstream distributions for thermal, pick-up and accelerated
protons from the Monte Carlo simulation, and the SWICS and HI-SCALE
measurements (see Fig.~1 of \cite{Gloeck94}) taken in the frame of the
spacecraft on the downstream side of the Day 292, 1991 shock.  The solar
wind and pick-up proton parameters are fairly tightly specified, so that
the ratio of the particle mean free path to its gyroradius, \teq{\eta
=\lambda /r_g} is essentially the only free model parameter. The
efficiency of acceleration of low energy ions in oblique shocks, i.e.
the normalization of the non-thermal power-law, is sensitive
\cite{EBJ95,BOEF97} to \teq{\eta}, which was adjusted to obtain a
reasonable ``fit'' to the data. For the large angle scattering case, the
downstream fit in the left hand panel of Fig.~1 models the accelerated
protons well for a {\it rigidity-independent} \teq{\eta = 5.5\pm 1.5}, a
value consistent with a moderate level of field turbulence; the
uncertainty in \teq{\eta} is due mostly to the observational uncertainty
in the shock obliquity \teq{\theta_{Bn1}}. The non-thermal proton
distribution is composed virtually entirely of accelerated pick-up ions
that are energized much more efficiently than thermal solar wind
\teq{H^{+}} ions. Note that the distribution of accelerated \teq{He^+}
pick-up ions reported by \cite{Gloeck94} for this shock can be modeled
\cite{BS05,SB06} by the {\it same} scattering parameter \teq{\eta =5.5}.

The left hand panel of Figure~1 also displays results for the field-line
wandering implementation as prescribed in
Eq.~(\ref{eq:theta_eta_momdep}), for the specific case of
\teq{\eta_0=30} and \teq{\theta_0=0.74}.  This case approximates a
\teq{\kappa_{\perp}\sim 10^{-4}\kappa_{\parallel}} scenario, i.e. a
variance of \teq{\sigma^2\sim 0.03 B^2} for the GJ99 turbulence
analysis.  The key conclusion is that the FLW model underpredicts the
Ulysses data downstream, largely because the retention of ions in the
shock layer is basically less efficient than in the LAS diffusion
picture. Trial runs with higher values of \teq{\eta_0} and
\teq{\theta_0} improve this situation somewhat, but not sufficiently to
match the observations.  This is a generic difficulty for the FLW
implementation, that can be truly rectified only for larger ratios
\teq{\kappa_{\perp}/\kappa_{\parallel}} and variances \teq{\sigma^2}. 
Such turbulent conditions are more commensurate with conditions in IP
shocks than the low variance constructs that were employed in GJ99 that
are appropriate to the quiet interplanetary solar wind. A fuller
exploration of such turbulence parameter space will be performed in
future work.

As highlighted in \cite{SB06}, an informative diagnostic on the
acceleration model is to probe the spatial scale of diffusion upstream
of the shock.  Results are illustrated in the left hand panel of Fig.~1
via the display of upstream distributions of high energy particles at
different times, i.e. distances from the shock. These exhibit a
characteristic ``peel-off'' effect \cite{Lee82} where superthermal ions
become depleted at successively high energies the further the detection
plane is upstream.  Fluxes for two different \teq{H^+} ion energies, 5
keV and 200 keV, were obtained from spectra like those in the left hand
panel in the Fig.~1, and are displayed in the right hand panel of the
Figure, together with corresponding data from Fig.~3 of \cite{Gloeck94}
for identical energy windows.  The normalization of the Ulysses data was
established by averaging over 3 day intervals, whereas the model
normalization was adjusted to match observed fluxes around half a day
downstream of the shock.

As particles diffuse upstream of the shock against the convective flow,
high energy ions with a mean free path \teq{\lambda\propto r_g^{\alpha}}
establish an exponential dilution in space/time upstream. For the LAS
case with \teq{\alpha=1}, this spatial scale of the exponential decline
is more or less identical to that of the model (modulo plasma
fluctuations) for the same choice of \teq{\eta=5.5} that optimized the
spectral agreement. This attractive concordance was featured in
\cite{SB06}, and is highly suggestive that ion diffusion in strong field
turbulence that is intimately connected to the shock is an integral part
of the acceleration process.  In contrast, Fig.~1 indicates that the
diffusive scale upstream for the \teq{\alpha =1/3}, \teq{\eta_0=30} FLW
model is far too short for 200 keV ions to accommodate the data.  This
principally arises because the dominant contribution to the diffusion
length is \teq{\ell\sim \kappa_{\parallel}\sin^2\theta_{Bn1}/u_1}, which
is much smaller at 200 keV energies than in the LAS case, by virtue of
the form in Eq.~(\ref{eq:theta_eta_momdep}).  Preliminary runs indicate
that, for the FLW construction, while \teq{\ell} can be increased to
match the observations, it is more difficult to model the spectra than
in the LAS scenarios that describe highly turbulent \teq{\langle \delta
B/B\rangle \gtrsim 0.1} regimes. While the LAS simulation results are
consistent with the observed results, it is impossible to draw more
definitive conclusions discriminating between LAS and FLW models,
without a focus on ions of intermediate energy, say around 50--100 keV,
the subject of future work.

\section{Conclusions}
 \label{sec:conclusion}

This paper has modeled measurements of the phase space distributions for
protons observed by the Ulysses instruments SWICS and HI-SCALE in the
Day 292, 1991 shock using a Monte Carlo simulation of diffusive shock
acceleration. Two diffusion models were employed, large angle scattering
appropriate for highly turbulent regions, and one approximating the
characteristics of field line wandering as determined by Giacalone and
Jokipii \cite{GJ99}. In both, the injection of pick-up protons dominates
that of solar wind protons in this highly oblique shock. It was found
that while there was consistency between the LAS model and the data for
energetic protons above speeds around 600 km/sec, outlined at length in
\cite{SB06}, the FLW scenario explored here has some difficulty in
generating superthermal \teq{H^+} in the power-law domain with the
requisite efficiency.  This is attributed to the comparative lack of
cross-field diffusion (\teq{\kappa_\perp /\kappa_\parallel \lesssim
10^{-3}}) in the FLW model, which is implemented for variances
\teq{\sigma^2\lesssim 0.1 B^2} commensurate with field data for the
quiet, interplanetary solar wind.  Because a decline of
\teq{\lambda/r_g} with ion momentum is implicit in the FLW model,
upstream ramp diffusion scales are accordingly very short at 200 keV
energies, much shorter than the observed pre-shock scale in HI-SCALE
data, unless large values of \teq{\eta_0 \equiv\lambda_0/r_{g0}\gtrsim
130} for suprathermal ions are adopted.  Summerlin \& Baring \cite{SB06}
observed that the LAS model could consistently explain this upstream
ramp.  Investigation of the turbulence parameter space in the FLW
implementation is ongoing, to assess whether it absolutely requires
higher \teq{\kappa_\perp /\kappa_\parallel} in order to successfully
model the data.  However, the fact that the diffusive LAS model works so
well in coupling the spectral and spatial properties suggests that
diffusion in highly turbulent fields is an integral aspect of the
acceleration process at this shock.

\bibliographystyle{aipproc}

\end{document}